# Annihilative fractals of coffee on milk formed by Rayleigh–Taylor instability

Michiko Shimokawa and Shonosuke Ohta

*Department of Earth System Science and Technology, Kyushu University*

*Kasuga, Fukuoka 816-8580, Japan*

Abstract

We have discovered a new fractal pattern, coffee exhibits on a horizontal surface, with a fractal dimension $1.88 \pm 0.06$, when a heavy coffee droplet interacts with the surface of lighter milk. Parts of coffee pattern on the surface disappear due to Rayleigh–Taylor instability, and the annihilative behavior produces the fractal pattern. Coffee fractals and the Sierpinski carpet have common features, such as the fractal dimension, annihilative behavior, and an exponential decrease of the pattern area. It is suggested the formation of fractal follows a rule similar to that of the Sierpinski carpet.

PACS number: 89.75.Kd, 05.65.+b, 47.53.+n

Keywords: pattern formation, fractal, Rayleigh–Taylor instability, Sierpinski carpet

Putting a heavier fluid on a lighter fluid results in instability at the boundary between the two fluids, causing a disturbance called the Rayleigh–Taylor (RT) instability [1]. RT instability occurs in many situations, such as the overturn of the outer portion of the collapsed core of a massive star [2], the formation of high luminosity twin-exhaust jets in rotating gas clouds in an external gravitational potential [3], laser implosion of deuterium-tritium fusion targets [4], and electromagnetic implosion of a metal liner [5], which has been a fascinating topic for both science and technological applications. The boundary is disturbed by competition between surface tension and gravity [1]. In most experiments and models, disturbance with only one wave mode has been observed at the boundary [6-9]. Recently, a model with nonlinear surface tension was calculated whose disturbance has several wave modes [10]. In our experiments of RT instability, however, a fractal pattern without the characteristic length appears at the surface, as shown in Figs. 1 (b) and (c), which shows the disturbance distribution at the boundary. Ordinarily theories cannot predict this phenomenon. It is surprising that the fractal structure at the horizontal surface forms spontaneously. The fractal pattern is derived from the annihilative behavior of RT instability. This is the first study to report fractal annihilation in experiments. Here, we report on the experimental results of the annihilative fractal.

Coffee and milk are used in our experiments, which serve as the heavier and lighter fluids in RT instability, respectively. The coffee is made from instant coffee granules (Nescafe GOLDBLEND), enabling us to alter its density. Chinese ink is added to the coffee granules for image binarization and clear visualization. Milk provides a clear background for surface pattern observations. Coffee granules (2.0 g) and Chinese ink (3.0 ml) are poured into a beaker and mixed by placing the beaker in water at 100°C. The densities of the mixed coffee solution and milk are 1.71 and 1.03 g/ml, respectively. Milk is poured into a tall beaker until it reaches the mark of 3.5 cm, as shown in Fig. 1 (a). The beaker stands for 10 min or more for the fluid to come to rest. The milk is at room temperature: $17.0 \pm 3.0$°C. A 0.2 ml droplet of coffee solution is placed at the center of the milk's surface with a pipette, as shown in Fig. 1 (a). The time evolution of surface patterns is recorded using a digital video camera (Sony HDR-FX1) from the position of Camera 1 in Fig. 1 (a). The same digital video camera is used in the vertical observation, positioned at Camera 2 in Fig. 1 (a). A macro lens (Raynox DCR-150) is attached to the digital video camera to increase clarity.

The pattern in Fig. 1 (b) is observed on the surface about 70 s after a coffee droplet is placed on the milk's surface. The white base in Fig. 1 (b) is milk, and the black lines are exhibited by solution of coffee and Chinese ink. Patterns are investigated with a density–density correlation function and a box counting method in order to determine whether the pattern is a fractal. In these fractal analyses, image processing results are used as shown in the inset of Fig. 2 (b), whose image size is $640 \times 480$ pixels.

The density–density correlation function $C(r)$ is defined by $C(r) = <\rho(r+r') \cdot \rho(r')>/N$, where $N = \sum \rho(r)$. The density $\rho(r)$ at the position $r$ of a real scale is 1, if any part of the coffee pattern is on the pixel, otherwise $\rho(r)$ is 0. The average of $C(r)$ over five experiments is shown in Fig. 2 (a) on a log–log plot. If the disturbance of RT instability has characteristic wave modes, according to previous models [6, 7], $C(r)$ should show some peaks. $C(r)$, however, is almost logarithmic from 0.1 to 1.2 cm, which implies that the coffee pattern is a fractal without the characteristic wave mode [11, 12]. The slope of the logarithmic part yields exponent $A$. The fractal dimension $D$ is given by $D = 2 + A$. In Fig. 2 (a), $A = -0.14$ and $D = 1.86 \pm 0.06$. Figure 2 (b) shows the result by box counting [12]. Box sizes and numbers are represented as $\varepsilon$ and $N(\varepsilon)$. The fractal patterns yield the relationship $N(\varepsilon) \sim \varepsilon^{-D}$. Data in Fig. 2 (b) exhibit a logarithmic fit between 0.05 and 0.8 cm, which supports the assertion that the coffee pattern is a fractal. The fit of Fig. 2 (b) yields a fractal dimension $D$ equal to $-B = 1.86 \pm 0.07$. This value agrees with $D = 1.86 \pm 0.06$ obtained from the correlation function within error.

Next, we consider whether a fractal structure is also observed in the same experiments using other liquids. Therefore, we investigate the surface pattern formed by a magnetic liquid instead of a coffee solution. The obtained pattern is also a fractal with the dimension $1.91 \pm 0.03$, as shown in Fig. 1 (c). Although this structure seems orderly and the dimension is slightly large compared to the coffee fractals, we equate this pattern with the one for coffee. It is generally expected that patterns formed by RT instabilities are fractals.

Let us focus on the formation process of the coffee fractal. The formation was investigated using a digital video camera. The top images in Figs. 3 (a)–(c) show surface patterns captured using Camera 1 at $t = 10$, 30, and 90 s after coffee droplet is placed on the surface. The bottom images are captured from the side using Camera 2 at $t = 10$, 30, and 90 s, which correspond to the times for the top images. In the vertical observation, water is used instead of milk for better visualization. This substitution, most likely, has a negligible effect, since the densities of water and milk are similar.

The results of the formation process are as follows: As soon as the coffee droplet is placed on the surface, the coffee solution spreads. After the spread, color variations appear on the surface at $t =$ 10 s, as shown in the top image in Fig. 3 (a). The dark regions correspond to high coffee density and light regions to low density. The disturbance develops from the dark regions, shown in the bottom image of Fig. 3 (a). At $t =$ 20 s, the color variation becomes wide, as shown in Fig. 3 (b), in comparison with Fig. 3 (a). Then, plumes develop under the surface. This is a characteristic feature of RT instabilities. Once gravity pulls the denser solution downwards, the surface coffee solution continues to go down. As a result, the fractal pattern in Fig. 1 (b) is formed. The fractal pattern remains intact at 90 s, as shown in Fig. 3 (c). The bottom image of Fig. 3 (c) shows the coffee solution flowing toward the center of the pattern, resulting in the formation of stick-like patterns surrounding the fractal pattern. The pattern thins out with time and disappears about 7 min later.

This observation reveals that the fractal pattern is caused by the annihilative behavior of the coffee solution due to RT instability. Our coffee fractal is a new type of an annihilative fractal, not observed before. Ordinarily, fractal structures, such as river branches [11—13], bacteria colonies [11, 12, 14, 15], and crystallization [11, 12, 14, 16], continue to grow, increasing in area. Coffee fractals, however, are annihilative, and parts of the pattern gradually disappear due to RT instabilities. This is the first instance of observing fractal annihilation in experiments.

For understanding the formation mechanism of the annihilative fractal, we measured the quantifiable data. Figure 4 (a) shows the time dependence of the fractal dimensions $D$ for coffee densities of 2.04, 1.71, and 1.54 g/ml. The density–density correlation functions are then calculated for the measurements of $D$. The fractal analyses are applied only at $t > t_1$. $t_1$ depends on coffee solution densities, and it is 43, 70, and 90 s for 2.04, 1.71, and 1.54 g/ml, respectively. Fractal dimensions, however, are almost constant in time for various densities within error although the patterns change with time, as shown in Figs. 1 (b) and 3 (c). The average fractal dimension is $1.88 \pm 0.06$, which is shown as a solid line in Fig. 4 (a). Figure 4 (b) shows the surface pattern density $\sigma(t)$ at $t$ for a coffee solution density 1.71 g/ml. $\sigma(t)$ is obtained from the brightness of pictures, such as the top images in Fig. 3. If all pixels of the picture are covered with the coffee solution, $\sigma(t) = 1$. And $\lim_{t \to +\infty} \sigma(t) = 0$ at $t \to +\infty$, which means that the coffee solution vanishes. The behavior of $\sigma(t)$ at $t > t_2$ is different from that at $t < t_2$, where $t_2$ is 14.8 s in Fig. 4 (b). For $t < t_2$, $\sigma(t)$ increases with time since the coffee droplet continues to spread at the milk's surface. After $t_2$, $\sigma(t)$ decreases with $t$, where the coffee solution stops spreading and goes down vertically. In the region, the fractal pattern is observed, which demonstrates that the behavior at $t > t_2$ is essential for fractal formation. The decrease of $\sigma(t)$ reinforces the assertion that the coffee fractal is annihilative. The exponential function, shown in Fig. 4 (b) as the solid line, is the best fitting line at $t > t_2$. Therefore, we assume $\sigma(t) \sim \exp(-t/\tau)$, where the average time constant $\tau$ is $69.8 \pm 10.2$ s for the coffee density 1.71 g/ml. $\tau$ almost agrees with $t_1 = 70$ s when we apply fractal analysis

The results in Figs. 4 (a) and (b) are reminiscent of the Sierpinski carpet [11], which is a representative mathematical model of annihilative fractals, shown in the inset of Fig. 4 (a). The fractal dimension is $\log 8 / \log 3 = 1.89$, which almost agrees with $1.88 \pm 0.06$ of the coffee fractals. The square measure $S(n)$ is $(8/9)^n$ at $n$ steps, where $S(0) = 1$ at $n = 0$ steps. For $n = +\infty$, $\lim_{n \to +\infty} S(n) = 0$. This indicates annihilation of the pattern, which is also similar to the coffee fractals. In follows, we discuss the formation process of coffee fractals, compared with the Sierpinski carpet.

Here, let us focus on the behavior of the correlation function. The correlation function for the Sierpinski carpet with small steps has peaks. For a large step number, no peak is found. We have already found the same behavior for our coffee pattern. At $t < t_1$, several peaks are found in the correlation, and no peak is found at $t > t_1$. We can understand the behavior as follows. With small steps or a short time from the start of annihilation process, the processes have not repeated sufficiently and the pattern is still a pre-fractal, which has some periodic structure. Then, the correlation has several peaks. After sufficient annihilation processes, the pattern has enough self-similarity structure. Then, the correlation does not have a peak, and we can determine the fractal dimension. The consideration yields the understanding for a decrease of $t_1$ with coffee solution density as mentioned in Fig. 4 (a). It is easier for the solution with a high density to go down vertically than for the lighter density solution, which causes an early annihilation process for the higher density solution. Thus, $t_1$ decreases with density. We can understand that the sufficient annihilation process is important for the formation of fractal.

Next, let us consider the number of annihilation processes required to recognize the coffee fractals and the Sierpinski carpet as fractals. For the Sierpinski carpet, six or more steps are needed in our correlation analysis. Meanwhile, the coffee pattern becomes a fractal after $t_1$ s as mentioned before. Here, remember $t_1 \sim \tau$ for our coffee pattern, where $\tau$ gives the decay ratio of pattern density $\sigma(t)$. This leads to the consideration that the pattern density has decreased sufficiently at $\tau$, i.e., the annihilation process has performed sufficiently and the pattern can be recognized as a fractal. Hence, $\tau$ almost agrees with $t_1$. In the condition that $t$ of the coffee fractal corresponds to $n$ of the mathematical models, $\sigma(t) \sim \exp(-t/\tau)$ has a common rule with $S(n) \sim (8/9)^n$ of the Sierpinski carpet. The equation $(8/9)^n = 1/e$ gives a minimum step number $n$ for recognizing the pattern as a fractal, based on this assumption. Then, the solution shows $n \sim 8$. This almost agrees with six, obtained from our correlation analysis for the Sierpinski carpet.

It is surprising that so many common features exist between coffee fractals and the Sierpinski carpet, such as (i) fractal dimension, (ii) annihilative behavior, (iii) an exponential decrease of the pattern area, and (iv) agreement of $n$. We expect that a rule similar to that for Sierpinski carpet exists for the formation of coffee fractals.

The pattern of the Sierpinski carpet, however, differs from that of the coffee fractal. In future, one should certainly compose a new annihilative model based on the Sierpinski carpet, which shows a similar pattern to the coffee fractal in satisfying these features common to both the Sierpinski carpet and coffee fractals. The proposal of a new model would lead to more advanced understanding, such as the effects of convection and turbulence on coffee fractal formation.

In summary, we have discovered an annihilative fractal pattern in an experiment of RT instabilities, using coffee droplets and milk. The fractal dimension is 1.88, which is constant in time and for varying densities of coffee solution. Side-view observations show that the annihilative behavior of the surface solution is essential for the fractal formation. The pattern area decreases with time according the exponential function. We discussed the formation process of the coffee fractal and comparing it to the Sierpinski carpet. The coffee fractal and Sierpinski carpet share common features, such as (i) fractal dimension, (ii) annihilative behavior, (iii) an exponential decrease of the pattern area, and (iv) agreement of $n$. This suggests that the Sierpinski carpet is a good model for explaining the coffee fractal formation process. The fractal structure must be formed according to a rule similar to that for the formation of Sierpinski carpet.

We thank H. Honji for proposing the experiment with the magnetic fluid and for fruitful discussions. We are also grateful to Y. Suguhara, Y. Takahashi, N. Mitarai and H. Nakanishi for helpful suggestions; M. Homma, K. Kishinawa and A. Saeki for discussions; and H. Honjo, H. Sakaguchi and H. Katsuragi for comments. This study is supported by the Japan Society for the Promotion of Science for Young Scientists.

References

1. D. H. Sharp, Physica D **12**, 3 (1984).

2. L. Smarr, J. R. Wilson, R. T. Barton and R. L. Bowers, Ap. J. **246**, 515 (1981).

3. M. L. Norman, L. Smarr, J. R. Wilson and M. D. Smith, Ap. J. **247**, 52 (1981).

4. M. H. Emery, J. H. Gardner and J. P. Boris, Phys. Rev. Lett. **48**, 677 (1982).

5. R. A. Gerwin and R. C. Malone, Nucl. Fusion **19**, 166 (1979).

6. U. Alon, J. Hecht, D. Ofer, and D. Shvarts, Phys. Rev. Lett. **74**, 534 (1995).

7. Y. Srebro, Y. Elbaz, O. Sadot, L. Arazi, and D. Shvarts, Laser and Particle Beams **21**, 347 (2003).

8. P. Brunet, G. Gauthier, L. Limat, and D. Vallet, Experiments in Fluids **37**, 645 (2004).

9. Vinningland, J. L. Johnsen, O. Flekkoy, E. G. Toussaint, R. and Maloy, K. J. Phys. Rev. Lett. **99**, 048001 (2007).

10. J. Garnier, C. Cherfils-Clerouin and P.-A. Holstein, Phys. Rev. E **68**, 036401 (2003).

11. B. B. Mandelbrot, *The Fractal Geometry of Nature*. Freeman, San Francisco (1982).

12. K. Glass and L. Glass, *Understanding Nonlinear Dynamics*, Springer-Verlag (1995).

13. I. Rodriguez-Iturbe and A. Rinaldo, *Fractal River Basins: Chance and Self-Organization*, Cambridge University Press (1997).

14. P. Ball, *The Self-Made Tapestry*, Oxford University Press (2001).

15. T. Matsuyama, and M. Matsushita, Crit. Rev. Microbiol. **19**, 117 (1993).

16.  M. Matsushita, M. Sano, Y. Hayakawa, H. Honjo and Y. Sawada, Phys. Rev. Lett. **53**, 286 (1984).

Figure captions

Fig. 1 (a) Schematic drawing of the experimental procedure. (b) Coffee fractal pattern obtained from a coffee solution density of 1.71 g/ml 70 s after a droplet is left on the surface of milk. (c) Fractal pattern formed by magnetic fluids instead of coffee solution. Patterns (b) and (c) were captured using Camera 1 in (a).

Fig. 2 Fractal dimensions calculated from (a) the density–density correlation function and (b) the box counting function. The inset of (b) is a binary image of Fig. 1 (b). The distance between two pixels, correlation function, box sizes and box numbers are represented as $r$, $C(r)$, $\varepsilon$ and $N(\varepsilon)$, respectively. The slopes of the data are $A = -0.14$ and $B = -1.86$. Fractal dimensions obtained from (a) and (b) are $D = 2 + A = 1.86$ and $D = -B = 1.86$, respectively.

Fig. 3 Formation process of coffee fractal patterns in experiment of coffee solution density 1.71 g/ml. The top and bottom images are captured using Cameras 1 and 2 of Fig. 1 (a). In the bottom images, water is used instead of milk for visualization. Inset in the bottom image of Fig. 3 (a) is the extended picture around the boundary between coffee solution and milk. Figures 3 (a), (b) and (c) are observed at 10, 20 and 90 s, respectively, after the coffee droplet is left on the surface.

Fig. 4 (a) Time dependence of fractal dimensions $D$ from the density–density correlation function. Data, obtained from experiments of coffee solution densities at 2.04, 1.71 and 1.54 g/ml, are shown as open circles, open squares and open triangles. The fractal analysis is possible from time $t = t_1 =$ 70 s after the coffee droplet is left on milk surface for the density 1.71 g/ml. The average fractal dimension over time and for different densities is $1.88 \pm 0.06$, which is drown as a solid line. (b) Fractal pattern density $\sigma(t)$ plotted against the time $t$ for a coffee solution density of 1.71 g/ml. The shape of $\sigma(t)$ changes at $t_2 \sim 14.8$ s. For $t > t_2$, $\sigma(t)$ can be fit with a exponent function $\sigma(t) \sim \exp(-t/\tau)$, which is shown as a solid line. The time constant $\tau$ is $69.8 \pm 10.2$ s. The inset image is a Sierpinski carpet, which is an annihilative mathematical model.

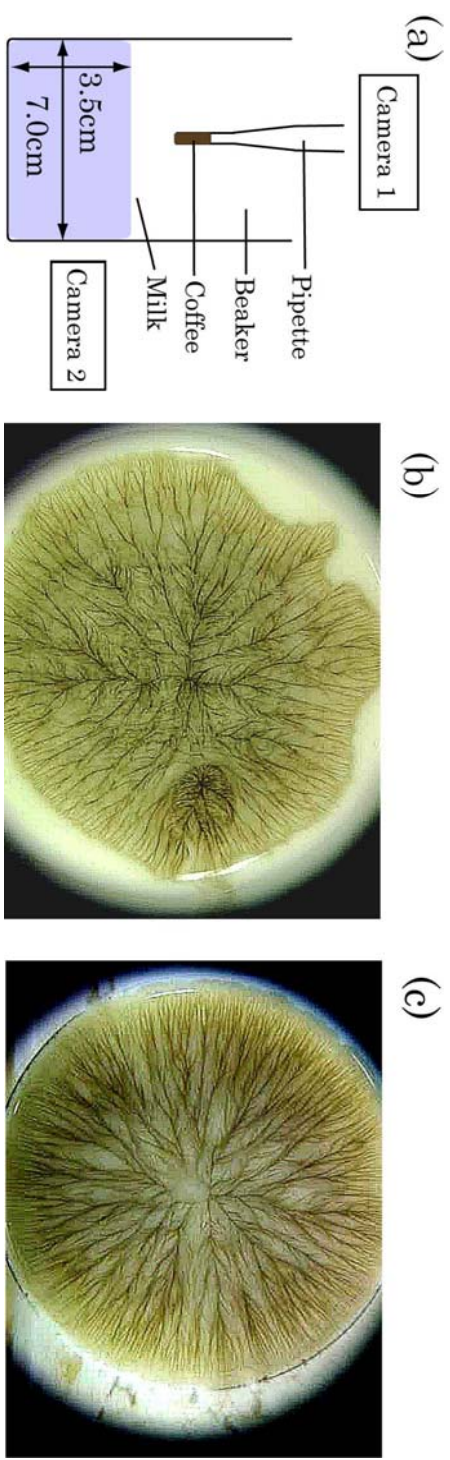

Fig. 1

Fig. 2

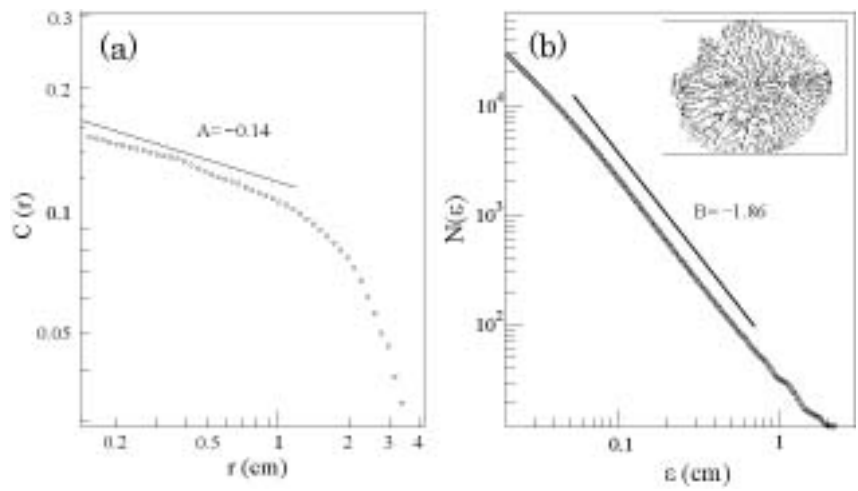

Fig. 3

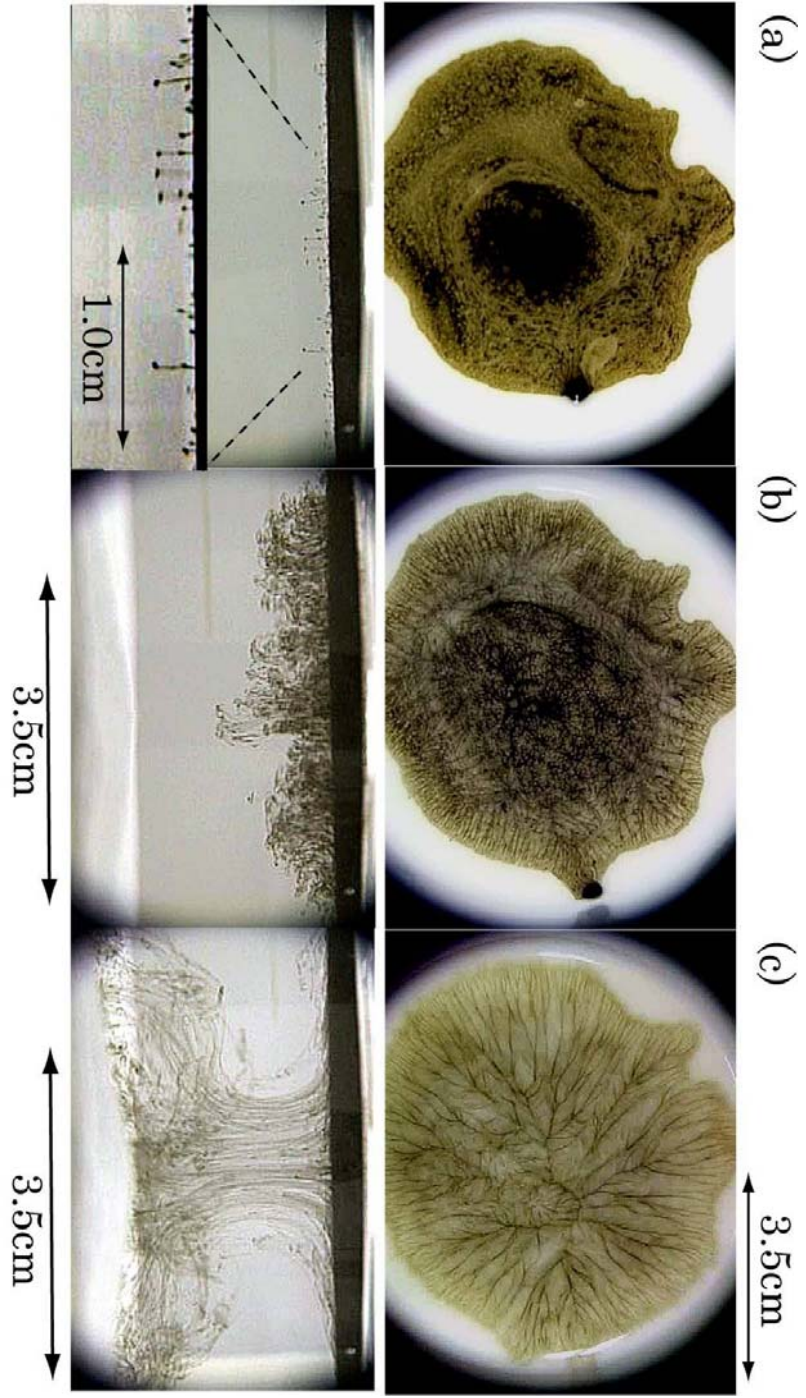

Fig. 4

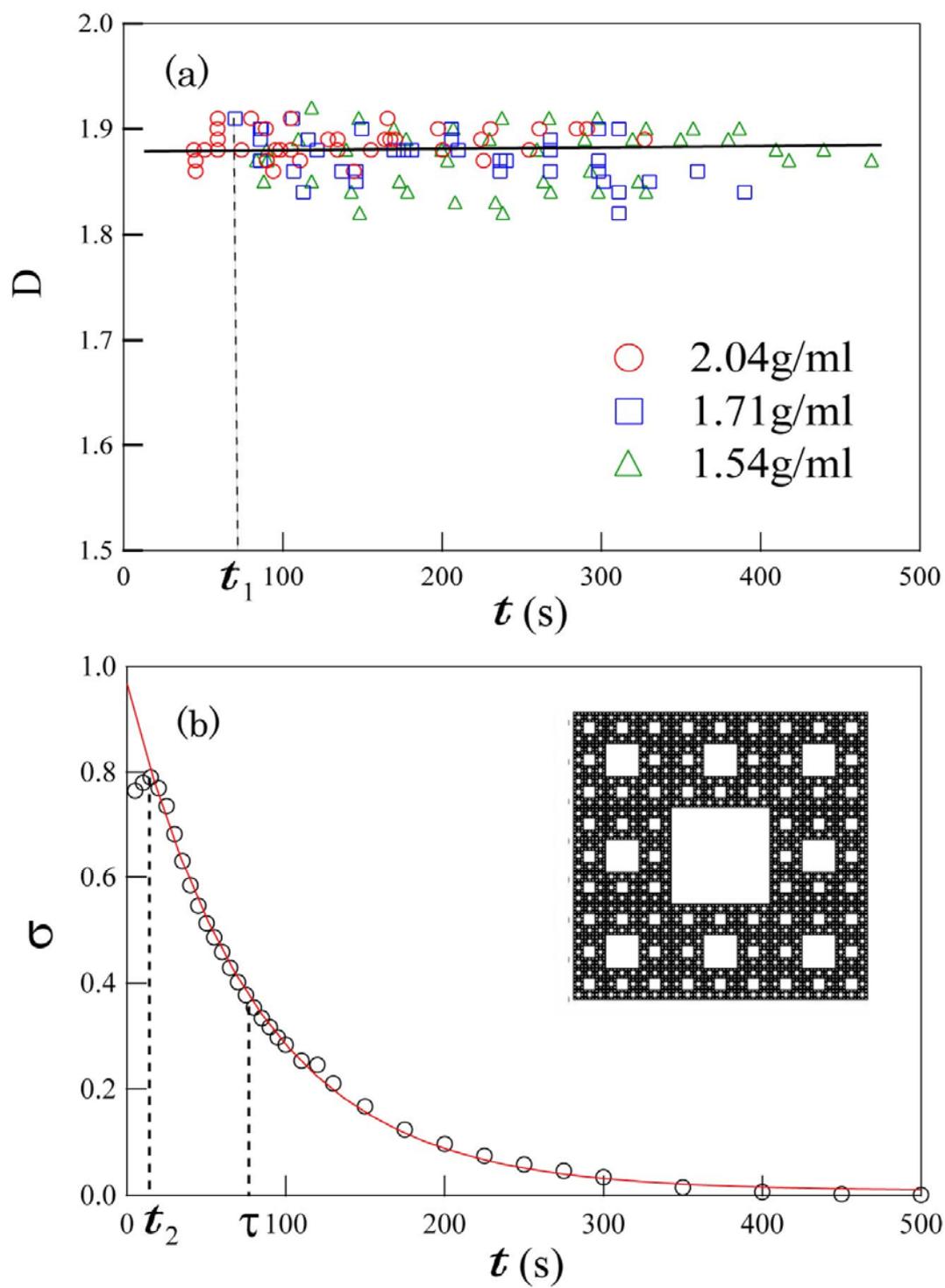